# Power-to-Gas as bridge between gas and electricity distribution systems: a sensitivity analysis of modeling approaches

Gabriele Fambri[a], Cesar Diaz-Londono[a], Andrea Mazza[a], Marco Badami[a], Robert Weiss[b]

[a] *Dipartimento Energia, Politecnico di Torino, Corso Duca degli Abruzzi 24, 10129 Torino, Italy*
[b] *VTT Technical Research Centre of Finland Ltd., P.O. Box 1000, 02044 VTT, Espoo, Finland*

ABSTRACT

*Power-to-Gas (P2G) has been one of the most frequently discussed technologies in the last few years. This technology allows producing $CO_2$ free fuels. Thanks to its high flexibility, it may offer services to the power system, fostering Variable Renewable Energy Sources (VRES) and the electricity demand match, mitigating the issues related to VRES overproduction. The role of P2G plants connected to the transmission system as flexibility asset has been extensively analyzed in the literature. Conversely, the analysis of these systems used at distribution level has only been dealt with in a few studies: however, in this case critical operation conditions can easily arise, both on electrical and gas infrastructure. This article presents a methodological analysis on the impact of different simulation approaches when P2G is installed at distribution system level. The choice of the most appropriate modeling approaches for electricity and distribution grids is required in order to avoid overestimating or underestimating the potential flexibility that P2G plants connected to distribution networks can offer. The aim of this paper is to understand the impact of different modeling approaches in order to determine whether, and under which conditions, this is acceptable. An illustrative case study has been developed to perform this kind of analysis. The results demonstrated that it is important to take into account the electric distribution network topology, as the performance of P2G plants could be affected by their placement in the network. Neglecting the dynamics of a gas network or the interactions between P2G plant components under a low gas demand condition can lead to an underestimation of the flexibility of the entire system. If the demand for gas is high enough, the use of simplified assumptions that do not consider the dynamics of the gas network and P2G systems may be acceptable.*

KEYWORDS

*Renewable energy integration, Electricity distribution system, Gas distribution system, Multi-energy system, Power-to-Gas, Modeling*

**ACRONYMS**

| | |
|---|---|
| BFS | Backward Forward Sweep |
| $CH_4$ | Methane |
| $CO_2$ | Carbon dioxide |
| DSO | Distribution System Operator |
| $H_2$ | Hydrogen |
| HP | High Pressure |
| HV | High Voltage |
| G2P | Gas-to-Power |
| LHV | Lower Heating Value |
| LPEN | Lumped Parameter Electricity Network |
| LPGN | Lumped Parameter Gas Network |
| LPP2G | Lumped Parameter P2G |
| MP | Medium Pressure |
| MV | Medium Voltage |
| OLTC | On-Load Tap Changer |
| P2G | Power-to-Gas |
| $P2H_2$ | Power-to-Hydrogen |
| P2X | Power-to-X |
| PEM | Polymer Electrolyte Membrane electrolyzer |
| PV | Photovoltaic plants |
| RPF | Reverse Power Flow |
| SNG | Synthetic Natural Gas |
| SoC | State of Charge |
| TR | Transformer |

TSO          Transmission System Operator
VRES         Variable Renewable Energy Sources
WT           Wind Turbines

# 1. Introduction

The average temperature of the world is about 0.8 °C higher than pre-industrial levels as a result of anthropogenic $CO_2$ emissions [1]. The European Union has defined the *Clean Energy for all Europeans* legislation package [2] with the intention of countering this dangerous trend, and renewable energies will play a fundamental role in this decarbonization process [3].In order to preserve the correct and safe operation of the entire electricity network system, energy generations and loads need to be instantaneously balanced and regulated. This delicate equilibrium may be undermined by Variable Renewable Energy Sources (VRES), which, due to their intrinsic high volatility, intermittency and low predictability nature, make it harder to achieve an electricity network balance [4]. New flexibility resources that make the most of renewable production and, at the same time, preserve the correct operation of the electricity system are needed to cope with these problems [5],[6]. Storage technologies, such as electric batteries [7], pumped hydro storage [8] or compressed-air energy storage [9], can offer a certain degree of flexibility, and allow the energy produced from renewable sources to be stored and dispatched in a flexible manner. Nevertheless, as also concluded in [10], in order to further increase the flexibility of an electricity system, it is necessary to review the paradigm of energy systems: in particular, by considering the entire system and not just the electricity sector, and in this way the global flexibility would increase significantly thanks to the exploitation of new synergies between different energy-intensive sectors [11],[12]. Various sources of flexibility that are outwith the pure electricity sector have been analyzed in the literature, such as: thermal regulation systems for buildings [13],[14], electric mobility [15], [16], heat pumps connected to a district heating system [17], district cooling networks [18] and demand response systems [19]. The coupling of an electricity network with a gas network has been debated a great deal in recent years. Thanks to Power-to-Gas (P2G) technologies, it is possible to convert electrical energy into gaseous fuels. The term P2G is used indiscriminately for both plants that produce pure hydrogen (through electrolyzers, supplied with electricity, which provide hydrogen through distilled water splitting) and for plants whose output is Synthetic Natural Gas (SNG). In the latter case, hydrogen, obtained from the electrolysis process, reacts with $CO_2$ to obtain methane ($CH_4$). Adding a methanation process means adding an extra step of energy conversion and this therefore induces an increase in energy losses (that is, a lower efficiency). Nonetheless, producing SNG instead of hydrogen still offers certain advantages: i) the volumetric energy density of methane is about *four* times higher than that of hydrogen, and this allows a more efficient transport of energy [20]; ii) injecting hydrogen into a gas network can cause hydrogen embrittlement problems, which may lead to cracks in the network pipes, while SNG can be injected without any particular restrictions [21]; iii) the use of hydrogen for domestic use, although feasible, is not as safe as natural gas, because of a higher risk of ignition [21]; iv) the production of SNG makes it possible to promote and therefore increase the value of $CO_2$ capture technologies [22]. This article only deals with the production of SNG: for this reason, from now on the term P2G will be used to indicate this type of technology, while the plants in which only hydrogen is produced will be indicated with the acronym $P2H_2$.

P2G plants connected to the transmission system have been widely analyzed: the P2G potential was evaluated in [23] for a regional scenario in Germany: the size of the considered P2G plants was optimized in order to minimize the levelized cost of electricity. The participation of an industrial P2G plant on the energy market and on the ancillary services market was analyzed in [24] and [25]. The role of P2G plants in a 2050 near zero carbon dioxide emission scenario was analyzed in [20]. In [26], the P2G technology was used to store VRES over-generation in a gas transmission system. It was concluded, in [27], that, thanks to the use of distributed resources, including P2G, it was possible to reduce renewable energy curtailment and, at the same time, increase social welfare. A 100% large, renewable, energy-based (wind and solar) city scenario was achieved in [28] by connecting electricity, gas and district heating networks through P2G plants.

As was concluded in [29], [30] and [32], the role of P2G connected to local distribution systems has only been analyzed in a few papers. Several small-scale P2G plants (300-700 kW of electrolyser) were used in [33], to absorb the surplus generation of PV in electricity distribution networks in a German region. In [34], the authors studied network voltage regulation using alkaline electroliers and an On-Load Tap Changer (OLTC); the same $P2H_2$ model was also used in [35] to optimize the size and allocation of the plants connected to a distribution electricity network, with the goal of tackling the increasing penetration of VRES. The utilization of the $P2H_2$ technology was also analyzed in [30]: an energy conversion system was used to absorb the over-generation caused by the installation of PV plants on the distribution network. In this case, the use of energy $P2H_2$ was evaluated and compared with an alternative network expansion solution. A real-time platform was presented in [31] to simulate distribution networks connected to P2G plants, and the paper also presented an analysis of a case study in which P2G systems operated as distributed resources to absorb the network RPF. The interaction between gas, electricity and district heating networks, enabled by P2G, was exploited in [36] in an attempt to reach the complete decarbonization of an energy system based on wind and solar energy. A techno-economic analysis of a complete P2G unit was carried out in [37]: the authors defined the optimum size of a plant to optimize the system from a technical and economic point of view. In [32] and [38], the authors analyzed the possibility of providing voltage regulation in an electricity distribution system using P2G and the Gas-to-Power (G2P) technology. A novel P2G model, based on real data, was presented in [29], and such a model, which was connected to an electricity distribution network with a high share of RES distributed generation, was analyzed.

Distributed flexible resources are gaining more and more importance for the control and regulation of power systems: as proof of this trend, the European Union is now working on fostering the participation of aggregate distributed sources on the markets for ancillary services by offering flexibility to Transmission System Operators (TSO) [39]. In the future, as concluded in the SmartNet project [40], this kind of flexibility service could also be used by Distribution System Operators (DSOs) to balance their distribution networks. The presence of multi-energy infrastructures (such as gas and electricity networks) in the same district may help handle VRES over-generation at a distribution level, by reducing the effects witnessed for the transmission system, and actively support the regulation of the overall electricity system.

The aim of this paper is to present a complete methodological analysis that will highlight the pros and cons of different simulation approaches. In fact, depending on the main modeling approaches presented in literature, one of the components that characterizes the multi-energy scenario is usually modeled in detail, while simplified models may be used for the other components. Table 1 shows, as an example, that none of the aforementioned studies on P2G in distribution systems analyzed the dynamics that take place between all three of the main components of a system, i.e., the gas network, the electricity network and the P2G plants. This paper is a progression of the work presented in [41], in which the interactions between an electricity grid, P2G plants and a gas network were analyzed through the use of complete models that were able to fully consider the possible interactions among all the elements. However, this simulation approach may lead to an increase in the computation time, and to the problem of accessing data that the owners may not disclose. On the basis of these issues, this study aims to understand what the effects of the most common simplifications that are made are, and to highlight whether, and in what cases, these simplifications could be acceptable. The modeling aspects analyzed in this paper are summarized as follows:

- The impact of taking into account the electricity distribution network topology. This aspect has already been analyzed in some literature contribution (e.g., [29]) and it needs to be included to conduct a complete analysis. Only in cases that deal with the determination of the potential of installed storage over a wide area may a simplified version of the High Voltage (HV) grid be considered (such as in [23], where the model only considers the equivalent capacity for different sub-regions, and in [43], where the potential of some flexible technologies was studied at a metropolitan level). However, if the aim of the study is that of establishing the impact of resources on the electrical system, its physical model should be included, otherwise such aspects as local over-generation cannot be recognized.
- The impact of gas network pressure dynamics. The demand for gas, when considering a high-pressure gas network. is normally high enough to be able to absorb the SNG generated by the surplus of renewables [26][42]. Accordingly, several studies have been carried out without considering any constraint related to the injection of gas into the gas network [23], [29], [43]. However, if the P2G plant is connected to a medium pressure (MP) distribution grid, the demand for natural gas could be very low, especially in the summer season, and may thus create a hurdle for SNG injection. Moreover, the utilization of a complete gas network model, which considers the gas flow and the pressure in the pipes, allows the *linepack* effect to be taken into account. The pipeline volume could be used as a vessel to store natural gas inside the network itself. This intrinsic flexibility allows a temporal mismatch to be obtained between gas injection and withdrawal. Whenever a P2G scenario is simulated, it is important that this flexibility is recognized in order to avoid underestimating the P2G flexibility potential [21].
- The interaction between the internal units of a P2G plant: the electrolyzer, the hydrogen buffer and the methanation unit. In several papers (e.g., [27], [32], [38], [42], [44]), the energy conversion process of a P2G plant was simplified by only taking into consideration the overall process efficiency, that is without considering the separate processes or the interactions between the internal components. This kind of approximation, as was also concluded in [26], may lead to an underestimation of the flexibility of a P2G unit. In fact, if all the components are considered as single units, the electrolyzer, whose load should vary in order to offer flexibility to the electricity network, is limited by the operation of the methanation unit, which in turn is bounded by the constraints of the gas network. On the other hand, when the components are considered separately, the operation of the electrolyzer results to be more flexible, because the hydrogen is not directly absorbed by the methanation unit and is instead accumulated in the buffer.

Table 1. Overview of previous studies on P2G in a distribution network scenario and the modeling assumptions.

| Ref. | Year | El. Network - physical model | Gas network - physical model | P2G - physical model |
|---|---|---|---|---|
| de Cerio Mendaza et al. [35] | 2015 | YES | NO | Only P2H$_2$ |
| Dalmau et al. [34] | 2015 | YES | NO | Only P2H$_2$ |
| Esterman et al. [33] | 2016 | NO | NO | NO |
| Khani et al. [38] | 2018 | YES | YES | NO |
| Robinius et al. [30] | 2018 | YES | NO | Only P2H$_2$ |
| El-Taweel et al. [32] | 2019 | YES | YES | NO |
| Salomone et al. [37] | 2019 | NO | NO | YES |
| Diaz-Londono et al. [31] | 2020 | YES | NO | NO |
| Mazza et al. [29] | 2020 | YES | NO | YES |
| Weiss et al. [36] | 2021 | NO | NO | YES |
| Fambri et al. [41] | 2021 | YES | YES | YES |

The remainder of this paper is presented as follows. Section 2 introduces the models used in this study; the analyzed case studies and the simulation input data are described in Section 3; Section 4 reports the results of the study, whereas Section 5 summarizes the conclusions.

## 2. The multi-energy system scenario

2.1. Electricity and gas networks

In order to carry out this kind of analysis, an illustrative multi-energy system scenario has been developed (see Figure 1). The case study has been presented in [41]. The multi-energy system includes an electricity Medium Voltage (MV) distribution network and an MP gas distribution network, coupled by means of P2G plants.

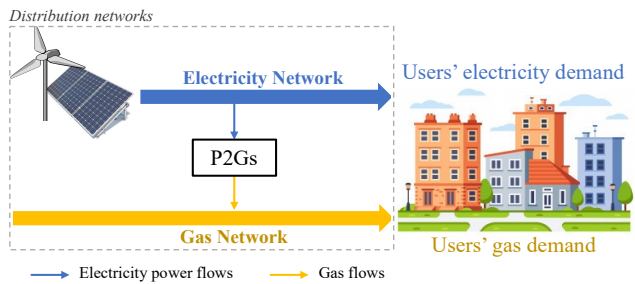

Figure 1. Multi-energy system scheme.

The data on the electricity and gas distribution network topologies were taken from real networks in northern Italy. The electricity distribution is related to an urban 22 kV-distribution system that operates in the City of Turin [45]: the network is composed of 43 electrical nodes, distributed over five feeders connected to the HV transmission system (220 kV) by means of three different HV/MV transformers (TRs) (see Figure 2a). The gas network topology was derived from [46]. The gas network is a medium pressure network of the 4th species, according to the DM 24/ 11/1984 1984 Italian classification [47] (operation pressure range: 1.5 – 5 $bar_g$). Considering the withdrawal nodes and the junction nodes, the network is overall composed of 70 nodes. The network has only one connection to the high pressure (HP) transmission system, i.e., the city-gate at node 1. Gas can flow, through the city-gate, from the HP to the MP portions, but not vice versa.

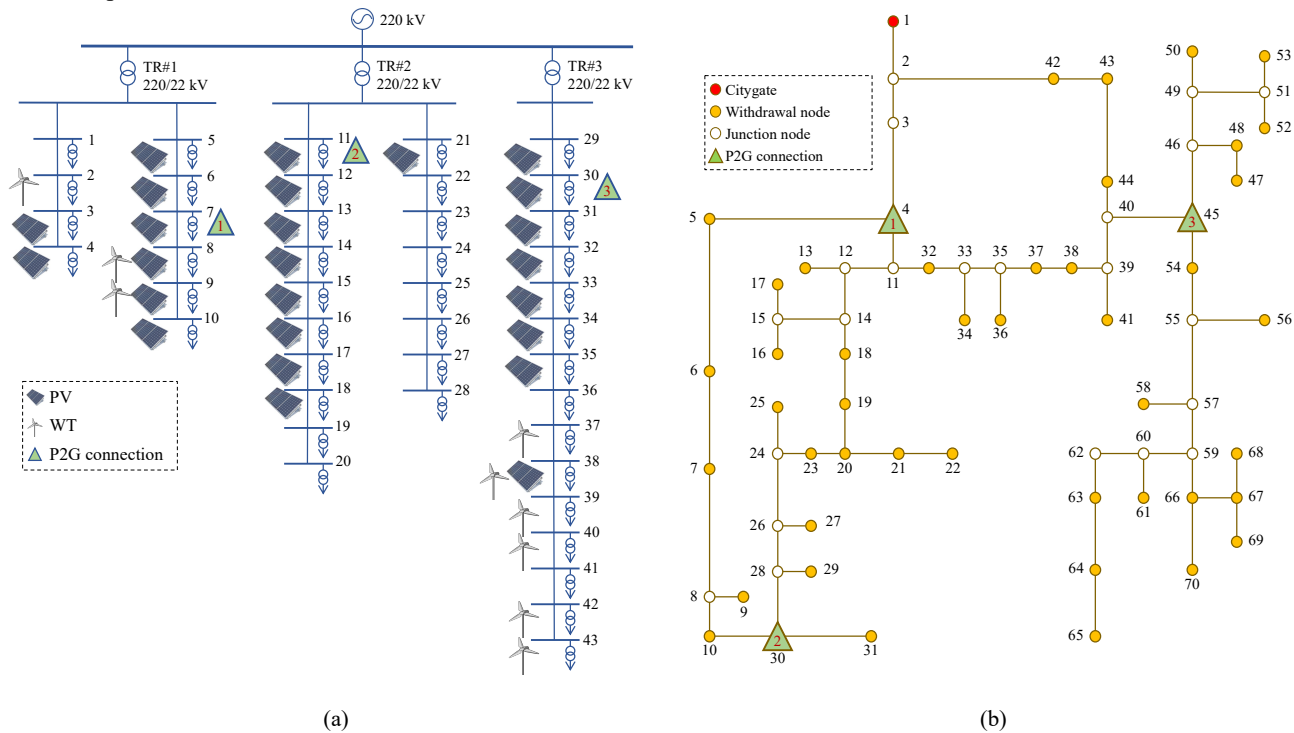

(a)          (b)

Figure 2. Networks topology and P2G connections. Electricity network (a). Gas network (b).

The scenario under analysis is characterized by a high number of installed VRES plants (photovoltaic plant, PV, and wind turbines, WT) connected to the electricity distribution grid. The scenario is assumed to cover the residential and tertiary sector electricity and gas users. The gas demand is mainly for domestic purposes and to heat buildings. The installed power, the peak power and the total yearly energy of VRES generation, the electricity demand and the gas demand are summarized in Table 2, while the total monthly energy demands and generations are reported in Figure 3. It is possible to note that, during hot months (from April to September), the VRES production is nearly two times more than the rest of the year, due to the higher solar irradiation, while the electricity demand is almost constant throughout the whole year, with a slight increase in summer, due to the activation of building cooling systems. The gas consumption is the most seasonal dependent: the gas demand during the coldest months is almost ten times higher than in summer, due to the high demand for building heating.

Table 2. Installed power, peak power and yearly energy of VRES, the electricity demand and the gas demand.

|  | Installed power [MW] | Peak power [MW] | Yearly energy [GWh] |
|---|---|---|---|
| PV | 14.3 | 10.8 | 21.3 |
| WT | 4.4 | 3.9 | 3.4 |
| El. demand | 12.3 | 6.0 | 29.7 |
| Gas demand | - | 23.0 | 36.0 |

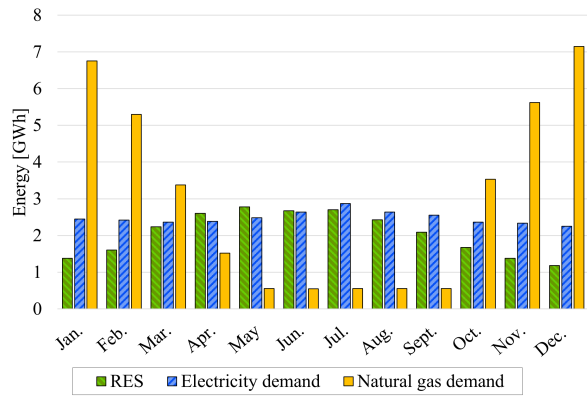

Figure 3. Monthly VRES generation, the electricity demand, and the gas demand.

These conditions generate two opposite situations in the hot months and in the cold months. In fact, the winter months are characterized by a relative low peak overproduction of renewable energy and a high demand for gas (see Figure 4c), while, in the summer months, during which the over-generation is higher, the demand for gas is considerably lower (see Figure 4b). The latter situation represents the most critical condition for the utilization of P2G flexible resources, because the large amount of VRES generation requires a higher exploitation of the P2G plants to balance the electricity network. However, the low gas demand could limit the SNG injections into the gas network.

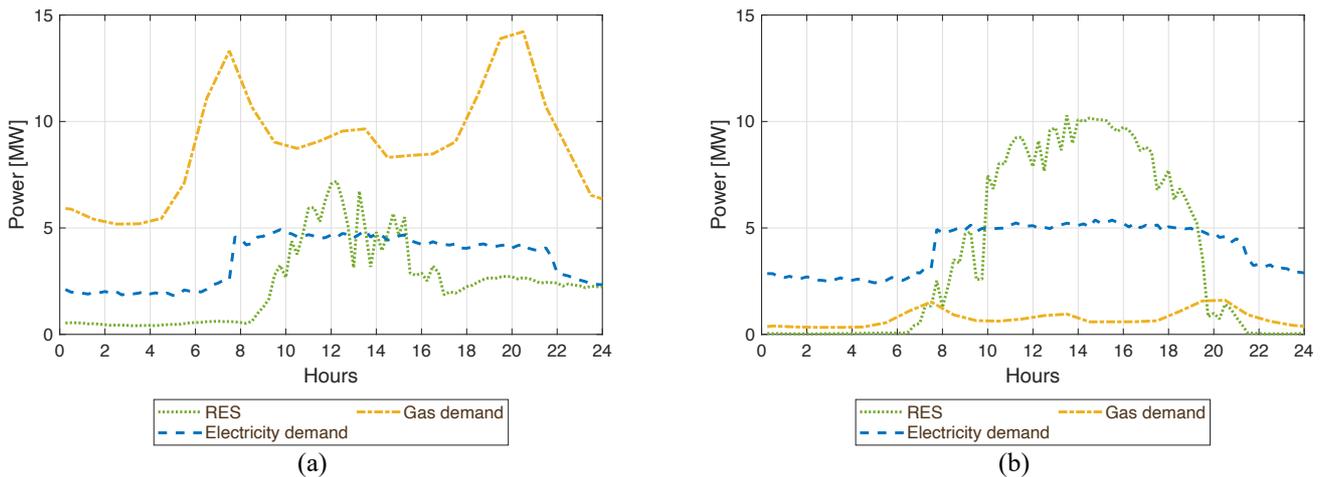

Figure 4. VRES generation, the electricity demand, the gas demand for a winter day (a) and the gas demand for a summer day (b).

As shown in Figure 2, three P2G plants are connected to electricity nodes 7, 11 and 30 and to gas nodes 4, 30 and 45. The three P2G plants have the same technical characteristics (Table 3): each plant is composed of a 1200 kW Polymer Electrolyte Membrane (PEM) electrolyzer (referred to the electrical input power), a hydrogen buffer capable of storing up to 3060 kWh of hydrogen (about 92 kg, considering the Lower Heating Value, LHV, of hydrogen, which is equal to 33.3 kWh/kg) and a methanation unit with a maximum SNG output equal to 43.2 kg/h (about 600 kW, referring to the $SNG_{LHV}$ output power).

Table 3. Size of the P2G plants and connections of the networks.

| Parameter | Unit | P2G#1 | P2G#2 | P2G#3 |
|---|---|---|---|---|
| Electrolyzer capacity | kW (el. input) | 1200 | 1200 | 1200 |
| Meth. unit capacity | kW (SNG output) | 600 | 600 | 600 |
| Hydrogen buffer capacity | kWh ($H_2$ LHV) | 3060 | 3060 | 3060 |
| El. network node connection | - | 7 | 11 | 30 |
| Gas network node connection | - | 4 | 30 | 45 |

2.2. P2G control in the distribution system

As depicted in Figure 2a, the electricity distribution network is connected to the high voltage transmission network by means of three HV/MV transformers connected with a busbar. When the network generation is higher than the consumption, the electricity flows from the distribution to the transmission system, thus causing unbalances in the transmission system that need to be regulated by the TSO. It could also happen that VRES over-generation only occurs in one portion of the network: in this case, the over-generation flows through the transformer, the connection busbar and the other transformers, and is then absorbed by the remaining distribution system feeders. Even though this situation does not affect the transmission system, the flow from MV to HV that occurs inside the transformer (the so-called Reverse Power Flow, RPF) needs to be avoided by the DSO, as the network protection systems are not able to guarantee a proper operation of the distribution system under such circumstances [48],[49]. It is worth noting that solving the RPF of the distribution system will also indirectly solve the issues at the HV interface, because the variability of the local generation is not captured by the TSO.

In this study, each P2G plant is connected to a feeder derived from a different HV/MV transformer (see Figure 2a), so that it can tackle all the local VRES over-generation that may exist within its feeder. In order to highlight the impact of local unbalancing, a critical scenario has been defined: as shown in Figure 5, about almost 60% of the energy over-generation occurs in the network portion derived from TR#3.

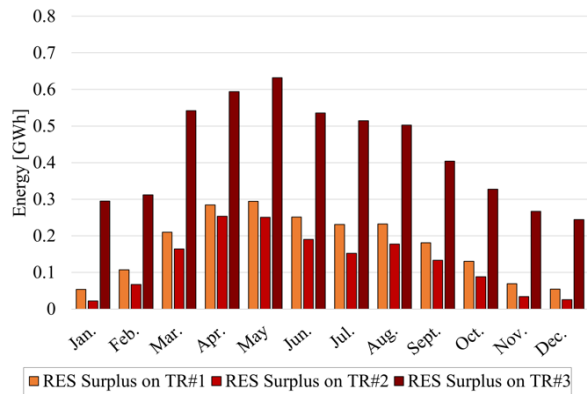

Figure 5. VRES over-generation (indicated in the figure as "Surplus") on the distribution network transformers.

When a local RES over-generation occurs, the PEM electrolyzer of the plant connected to the part of the network that is affected by over-generation is turned on. Thanks to their fast response and high flexibility, PEM electrolyzers are in fact able to follow the fast variation of VRES [50],[51]. The surplus electricity is converted into hydrogen and stored in the hydrogen buffer of the plant. The utilization of the electrolyzers is constrained by their nominal capacity and by the State of Charge (SoC) of the hydrogen buffer. The electrolyzers cannot operate at higher loads than their nominal capacity; moreover, if the hydrogen buffer of the plant reaches SoC 100%, the electrolyzer needs to limit its production, and thus cannot be used to absorb all the RPF. The cumulated hydrogen is used by the methanation unit for SNG production. The methanation unit is turned on when the SoC of the hydrogen buffer is higher than 50%. The SNG can be injected into the gas distribution network as long as the network pressure is below its upper limit (5 $bar_g$). If the pressure of the gas network is close to its upper limit, the production of one or two, or even of all the methanation units may to be blocked: in such a case, the methanation units that are already fully operational and which, as an additional criterion, have a larger amount of hydrogen stored in the buffer, are the first to operate.

### 3. Case studies

Different case studies are presented hereafter to analyze the value of modeling (i) the electricity topology, (ii) the gas pressure dynamics and (iii) the detailed P2G representation. All the aspects of the analyzed models are taken into account in the reference scenario. The same scenario was then simulated by neglecting a specific aspect in the modeling each time. The various scenarios thus obtained were compared with the reference one in order to identify the effects of various simulation approaches.

## 3.1. Reference case: complete models

The mathematical models of the reference case have been presented in more detail in [41].

### 3.1.1. Electricity grid

The electricity network is radial; under the hypothesis of the network being balanced (and this can be seen as a suitable approximation of the MV system), the power flow calculation, expressed in per unit, may be solved by applying the equivalent single-phase Backward Forward Sweep (BFS) algorithm [52]. The BFS considers $N$ load nodes and $B$ branches (with $N = B$ due to the radial network topology) and computes the branch current for every time step $t$ as follows:

$$\begin{cases} \underline{\mathbf{i}}_{B,t}^{(k)} = \mathbf{\Gamma}^T \cdot \underline{\mathbf{i}}_{N,t}^{(k)} = \mathbf{\Gamma}^T \cdot \left[ \underline{\mathbf{y}}_C \circ \underline{\mathbf{v}}_t^{(k-1)} + \underline{\mathbf{s}}_t^* \oslash \left( \underline{\mathbf{v}}^{(k-1)} \right)^* \right] \\ \underline{\mathbf{v}}_t^{(k)} = \underline{\mathbf{v}}_{1,t} - \mathbf{\Gamma} \cdot \underline{\mathbf{Z}}_B \cdot \underline{\mathbf{i}}_{B,t}^{(k)} \end{cases}, \forall\ t \in \mathbb{T} \quad (1)$$

where:

- $\underline{\mathbf{i}}_{B,t}^{(k)} \in \mathbb{C}^{B,1}$ represents the vector containing the complex currents at the time step $t$, calculated during the backward phase of the BFS method, at the $k$-th iteration,
- $\underline{\mathbf{v}}_t^{(k)} \in \mathbb{C}^{N,1}$ indicates the vector containing the complex voltages at the time step $t$, evaluated during the forward phase of the BFS method, at the $k$-th iteration,
- $\underline{\mathbf{i}}_{N,t}^{(k)} \in \mathbb{C}^{N,1}$ is the vector of the load nodes complex currents, at the $k$-th iteration;
- The inverse of the node-to-branch incidence matrix is the matrix $\mathbf{\Gamma} \in \mathbb{N}^{B,N}$. It provides the topological information about the network,
- The load admittances and the network admittance are included in the vector $\underline{\mathbf{y}}_C \in \mathbb{C}^{N,1}$,
- $\underline{\mathbf{s}}_t \in \mathbb{C}^{N,1}$ is the vector of the constant power loads, i.e., loads that absorb a constant power value at the time $t$ whatever is the value of the voltage at their nodes,
- $\underline{\mathbf{v}}_{1,t} \in \mathbb{C}^{N,1}$ is the vector containing the value of the slack node (indicated as 1 in Figure 6) at the time step $t$,
- $\underline{\mathbf{Z}}_B \in \mathbb{C}^{B,1}$ is a diagonal matrix collecting the values of the branch impedances.

Moreover, the mathematical operators $\circ$ and $\oslash$ are the Hadamard product and division, respectively; $k$ refers to the calculation iteration, $t$ to the time step and $\mathbb{T}$ is the set containing all the horizon time steps. The symbol * indicated the conjugate operation. Finally, the symbols $\mathbb{N}$ and $\mathbb{C}$ represent the set of natural and complex numbers, respectively

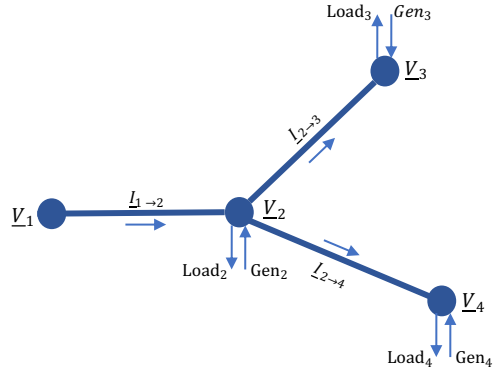

Figure 6. Scheme of the electricity network mathematical model.

### 3.1.2. Gas grid

The gas network model takes into account the gas flow in each pipe and the pressure in each node of the network (see Figure 7). The model is based on the Renouard equation for the mean pressure [53], the equation of state for ideal gases and the continuity equation. The mass flow between two nodes is determined from their difference in pressure through the Renouard relation: the gas flow between node $m$ and node $n$ is positive, if the pressure of node $m$ is higher than that of node $n$, and negative (flowing in the opposite direction) vice versa, i.e., the model is bi-directional. The pressure of each node is calculated using the equation of state as a function of the pressure and mass that exist in the node. The mass of gas in the node is given by the continuity equation, which considers the gas injections and withdrawals, and the gas flows that go from that node to the adjacent ones.

$$P_m - P_n = \left[ P_m - \sqrt{P_m^2 - 25.24 \cdot L_{m-n} \cdot \left( \frac{\dot{m}_{m-n} \cdot 3600}{\rho_{NG}} \right)^{1.82} \cdot D_{m-n}^{-4.82}} \right] \quad (2)$$

$$\frac{dP_m}{dt} = 10^5 \cdot \frac{R_{NG} \cdot T}{V_m} \cdot \dot{m}_m \quad (3)$$

$$\dot{m}_m = \dot{m}_{inj,m} - \dot{m}_{wit,m} + \sum_{i=1}^{N} \dot{m}_{m-i} \tag{4}$$

where:

- $P_m$ and $P_n$ are the pressures of nodes $m$ and $n$, respectively [bar];
- $L_{m-n}$ is the length of pipe $m-n$ [m];
- $\dot{m}_{m-n}$ is the natural gas flow inside pipe $m-n$ [kg/s];
- $\rho_{NG}$ is the natural gas density at standard conditions [kg/m³];
- $D_{m-n}$ is the diameter of pipe $m-n$ [mm];
- $V_m$ is the volume of the node [m³]
- $m_m$ is the mass [kg];
- $R_{NG}$ is the specific gas constant of natural gas [J/kg/K];
- $T$ is the natural gas temperature [K];
- $\dot{m}_{inj,m}$ and $\dot{m}_{wit,m}$ are the gas injection and withdrawal at node $m$, respectively [kg/s].

The model calculates the pressure in all the nodes of the network. This allows one to analyze how the volume of the gas network can be exploited to accumulate the gas inside the network, thanks to the linepack effect: storing gas inside the pipes increases the internal pressure of the network, and the accumulation can continue until the network pressure reaches the network operating pressure limit (in this case, 5 bar$_g$). If this feature were not taken into account, the injection of SNG into the network would be constrained by the instantaneous demand for gas, while, thanks to the physics of this system, the gas network allows a more flexible use of the methanation units.

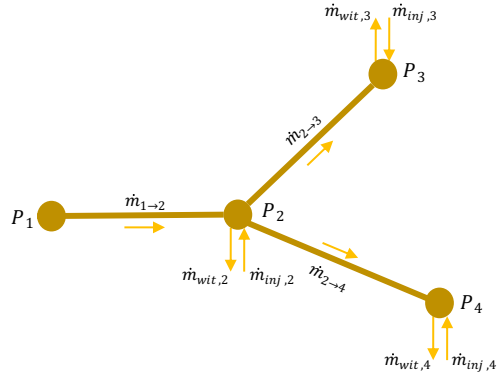

Figure 7. Scheme of the gas network mathematical model.

3.1.3. P2G plant

The main components of the P2G plant are the Polymer Electrolyte Membrane (PEM) electrolyzer, the hydrogen buffer and the methanation reactor (see Figure 8). The hydrogen buffer allows a decoupling to be made between the electrolyzer and the methanation unit. The produced hydrogen is accumulated in the buffer, and, in this way, the electrolyzer can work, even though, at that moment, the methanation unit does not use the produced hydrogen. Similarly, the methanation unit can use the hydrogen previously produced by the electrolyzer and operate independently. The electrolyzers are controlled to absorb the renewable over-generations that affect the transformers of the distribution network. The PEM electrolyzers are able to change their setpoint very quickly, within less than 2 seconds; since the simulation is performed with a time discretization step of 15 minutes, the ramp limit for the electrolyzers has been neglected. The produced hydrogen is accumulated inside the buffer: if the buffer reaches the maximum operating pressure (30 bar), the accumulation of hydrogen must be stopped. The methanation reactor is maintained in hot stand-by, and when the hydrogen buffer reaches a pressure of 15 bar, the unit is turned on. The upward and downward ramp rate constraints of the methanation reactor limit the load variation of these units: the maximum upward ramp rate is 3.8 kg/h, while the maximum downward ramp rate is 46 kg/h. As a result of technical limitations of this technology, the methanation unit cannot work below 50 % of its nominal capacity without being shut down. If the gas network reaches its maximum working pressure (5 bar$_g$), SNG generation must be limited to keep the gas network within the allowed pressure range.

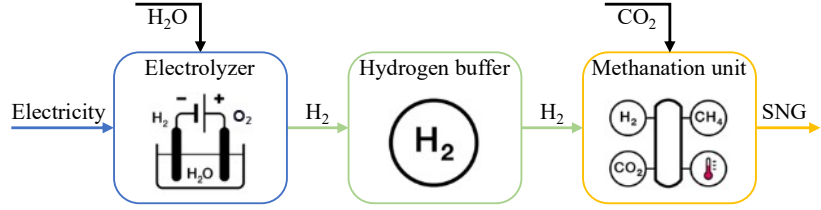

Figure 8. Scheme of the complete P2G model, adapted from [54].

3.2. The lumped parameter electricity network (LPEN) case

In the lumped parameter electricity network (LPEN) case, instead of simulating the whole electricity network, as in the Reference case, the electricity network is simplified by considering that all the loads and distributed generation are concentrated in a single node (see Figure 9). The RPF is calculated as the positive difference between the sum of the distributed generation in the network and the sum of all the loads. If the network distributed generation is lower than the network energy demand, the difference is assumed to be provided by the HV network:

$$RPF = \begin{cases} \sum gen - load, & if \sum gen - load > 0 \\ 0, & else \end{cases} \quad (5)$$

$$HV_{el} = \begin{cases} \sum load - gen, & if \sum load - gen > 0 \\ 0, & else \end{cases} \quad (6)$$

The three connections to the high voltage network (i.e., the three transformers) are not considered separately, but merged into a single connection point. The model is not able to identify any local RES over-generation that may affect the transformers of the distribution network, but only the total RES over-generation that affects the transmission system. Since it is not possible to identify in which part of the electricity grid the overproduction occurs, the excess energy is distributed equally over the three P2G plants.

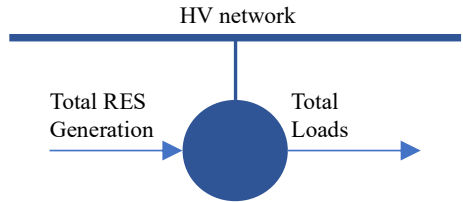

Figure 9. Schema of the electricity network lumped parameter model.

3.3. The lumped parameter gas network (LPGN) case

A simplified gas network model has been used for the lumped parameter gas network (LPGN) case (see Figure 10). The model does not take into account either the pressure evolution of the network nodes or the gas flow in each network pipe. All the users' gas withdrawals and SNG injections are considered to happen at the same points. If the gas demand is higher than the SNG injection, the difference is taken from the HP network:

$$Gas\ import = \begin{cases} \sum withdrawals - SNG\ injection, & if \sum withdrawals - SNG\ injection > 0 \\ 0, & else \end{cases} \quad (7)$$

The model cannot take into account the linepack effect: hence, SNG can be injected as long as it does not exceed the gas demand.

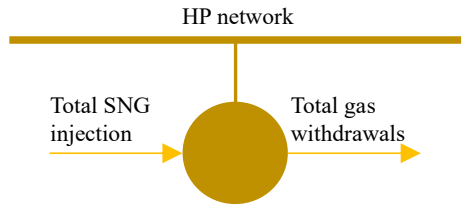

Figure 10. Schema of the gas network lumped parameter model.

3.4. The lumped parameter P2G (LPP2G) case

For this case, the P2G model does not consider the interaction between the main components of the plant. The entire process is summarized by means of fixed conversion efficiencies (see Figure 11). The electricity consumed by the plant is directly converted into SNG, without considering the internal dynamics of the plant, such as the hydrogen accumulation in the buffer, the methanation unit ramp rates or minimum load constraints. The electricity consumption and SNG generation are thus temporally linked to each other:

$$SNG = \eta_{P2G} \cdot (EL_{P2G} - EL_{aux}) \tag{8}$$

where:

- $SNG$ [kW] is the Synthetic Natural Gas production;
- $\eta_{P2G}$ is the efficiency of the entire energy conversion process;
- $EL_{P2G}$ [kW] is the electricity consumption of the plant;
- $EL_{aux}$ [kW] is the electricity consumption of the auxiliary components, which is considered to be constant and equal to 1% of the nominal electrical load of the plant.

For the sake of consistency, $\eta_{P2G}$ has been set equal to the average efficiencies over the full year, as simulated by the complete model.

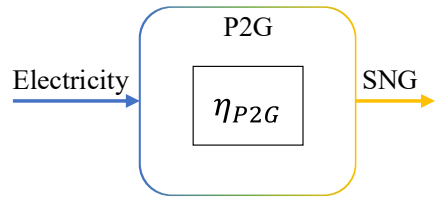

Figure 11. Scheme of the P2G lumped parameter model.

## 4. Results

In this section, the results of the three lumped parameter case simulations are compared with those of the Reference case by highlighting the main discrepancies and similarities. The Reference case was analyzed in depth in [41]; the reader may refer to [41] for details of the models.

4.1. Value of electricity network modeling

The electricity network lumped parameter model does not consider the different HV/MV connection points. Even though the overall network is balanced, in terms of energy generation and consumption, the optimum operation condition may not have been reached due to a local load/generation mismatch. For example, Figure 12 compares the energy balance of the whole electricity network in the Reference case versus the one simulated with the lumped parameter electricity network for a typical winter day. The LPEN case underestimates RPF: in fact, the P2G plants in LPEN appear to be able to absorb all the VRES over-generation, while the VRES over-generations in the Reference case are not totally absorbed, thus causing RPF (see the orange area in Figure 12a). Moreover, it seems that P2G#3 is activated in the Reference case, even when there is no need to actively absorb over-generation (i.e., during the nighttime). However, if the energy balance is computed at each HV/MV transformer level, local unbalances appear (see Figure 13)[1]. Since the lumped parameter model is not able to define where over-generation occurs, the activation of the P2G plants has not been properly coordinated. It could happen that a P2G is sometimes activated, even though there is no need to absorb local unbalances (see the balance of TR#2 in Figure 13) and that some local VRES over-generations are not detected by the lumped parameter model (see the balance of TR#3 in Figure 13). For these reasons, the optimal coordination of various P2G plants cannot be achieved and P2G#3 results to be used about 40% less frequently than in the Reference case, while P2G#2 is used about 50% more (see Table A. 1). It can be noted that, in the LPEN case, the P2G load (i.e., the sum of the electricity consumption of all three plants) is around 10% *lower* than in the Reference case, and this also affects the total SNG production of the plants, which leads to an underestimation of the injection of SNG into the gas network (see Table A. 3). This is due to the underestimation of the local RPF, whose sum results to be higher than the one that is seen by the transmission system. It is worth noting that the more different the local electricity imbalances are, the more marked the difference between the two models: in fact, the generation disparity within the network is greater in the winter months (in this period, about 65% of the over-generation takes place in the network portion derived from TR#3) and the difference in the use of P2G plants is about 20%. Unlike the winter months, the local grid mismatches are more homogeneous in the summer months, and the difference between the two modeling approaches is reduced to about 5%.

The RES over-generation, calculated with Eq. 5 (without considering the losses due to the Joule effect), corresponds to the network unbalance that affects the transmission system. Thus, even though the simplified model does not allow an optimum dispatchment of the P2G plant utilization or the evaluation of the best P2G plant network connection, it could be used for the high level qualitative evaluation of the P2G flexibility potential for transmission system balancing purposes (as shown in [37] and [43]).

---

[1] In order to evaluate the local unbalances in the LPEN case (which does not consider the HV/MV transformers), the reference model of the electricity network was run with P2G electricity load profiles obtained from the LPEN simulation.

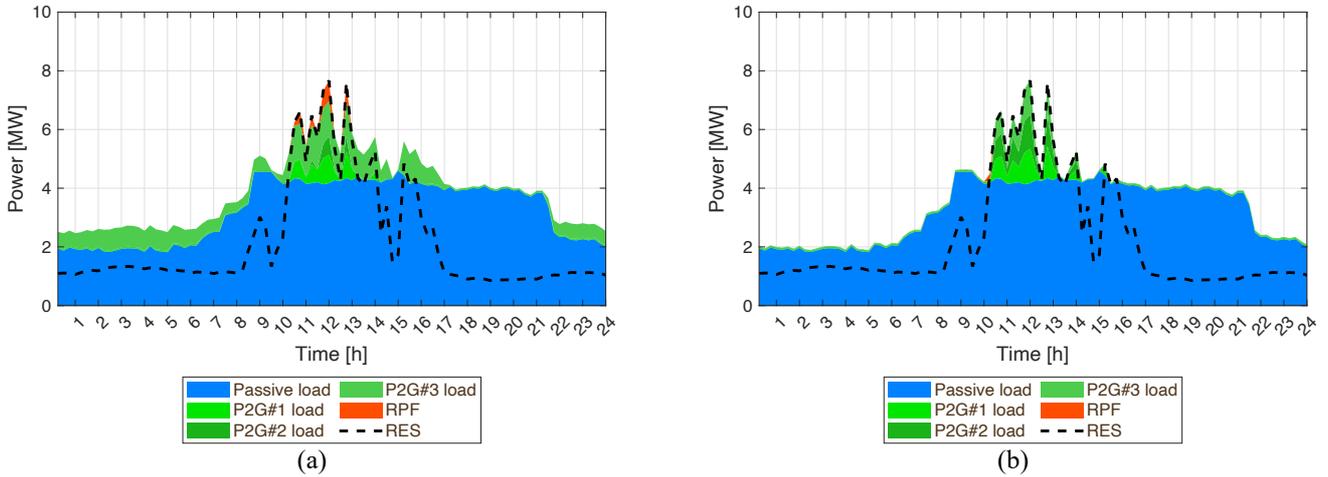

Figure 12. Electricity balance referring to the whole grid (winter day): (a) Reference case and (b) LPEN case.

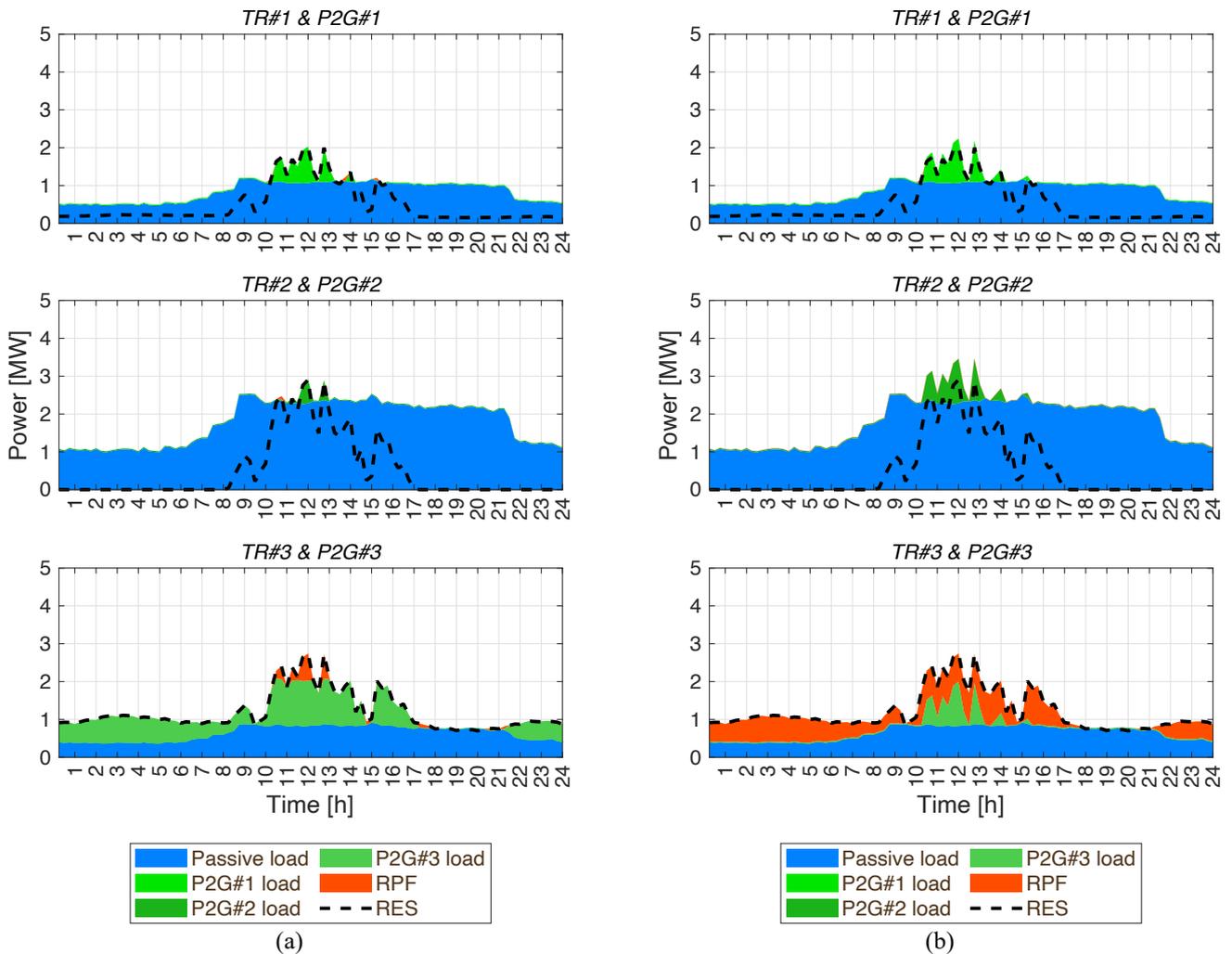

Figure 13. Electricity balance on HV/MV transformers (winter day): (a) Reference case and (b) LPEN case.

4.2. Value of gas network modeling

Unlike the reference model, the lumped parameter model does not take into account the pressure or gas flows in the gas network. Nevertheless, the results obtained with the lumped parameters model are almost the same as those of the Reference case, whenever only the heating season is considered (see Table A. 1-3). In these conditions, in fact, the production of SNG covers only 4% of the users' gas demand (see Table A. 3) and, at any timestep, the production of SNG is considerably lower than the users' gas demand (see Figure 14). All the injected SNG is directly consumed by the users and this behavior is clearly simulated by both models.

The demand for gas is highly seasonal, due to the use of gas to heat buildings: in the analyzed scenario, the gas demand during the hot season is about 10 times lower than in the winter season. In summer, the SNG production could exceed the users' gas demand (see Figure 15a)[2]; when this happens, the produced SNG could be stored by exploiting the gas network volume, thanks to the linepack effect. The accumulation of gas increases the pressure in the network (see the red curve in Figure 15a). When the maximum operating pressure in the network is reached, the SNG injection must be reduced (in the case reported in Figure 15a, the pressure reaches a level of 5 $bar_g$ at 15:15 and the P2G#1 and P2G#2 methanation units block their SNG injection to let the network lie within its operation pressure range). The accumulated SNG is used in the following hours to meet the gas demand (see the white areas in Figure 15a): when this happens, the gas stored inside the network decreases, as does the network pressure. The gas network lumped parameter model does not allow the network linepack to be taken into account: SNG injection is limited in order to always be lower than the network gas demand (see Figure 15b). Therefore, ignoring the intrinsic flexibility of the gas network limits the use of methane gas units, and, in the hot season, SNG production is underestimated by about 30% (see Table A. 3).

The limitation of the methanation units also affects the functioning of the electrolyzers: if the methanation units are unable to consume the hydrogen accumulated in the buffers, once the saturation of the hydrogen buffers is reached, the electrolyzers have to block their production of hydrogen and, consequently, they are no longer able to offer flexibility to the electricity grid. The underestimation of the flexibility of the gas network, induced as a result of the use of the simplified model, not only affects the gas network, but also the electricity network. In fact, the electrolyzers result to be less flexible, and to cause an overestimation of almost 190% of the RPF generated by the HV/MV transformers during the hot season (see Table A. 2 and Figure 16).

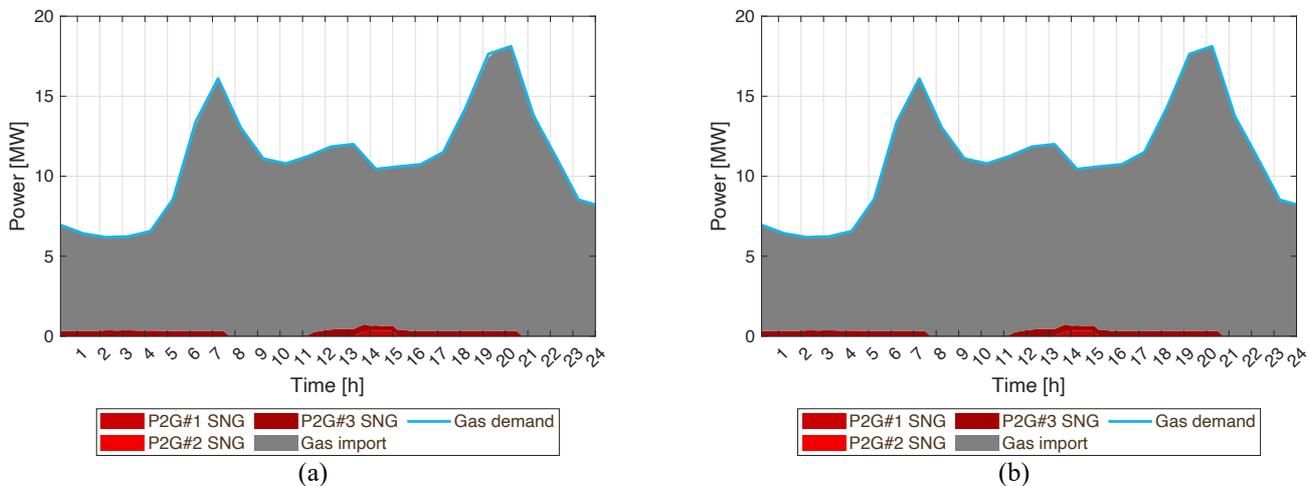

Figure 14. Gas network balance (winter day): (a) Reference case and (b) LPGN case.

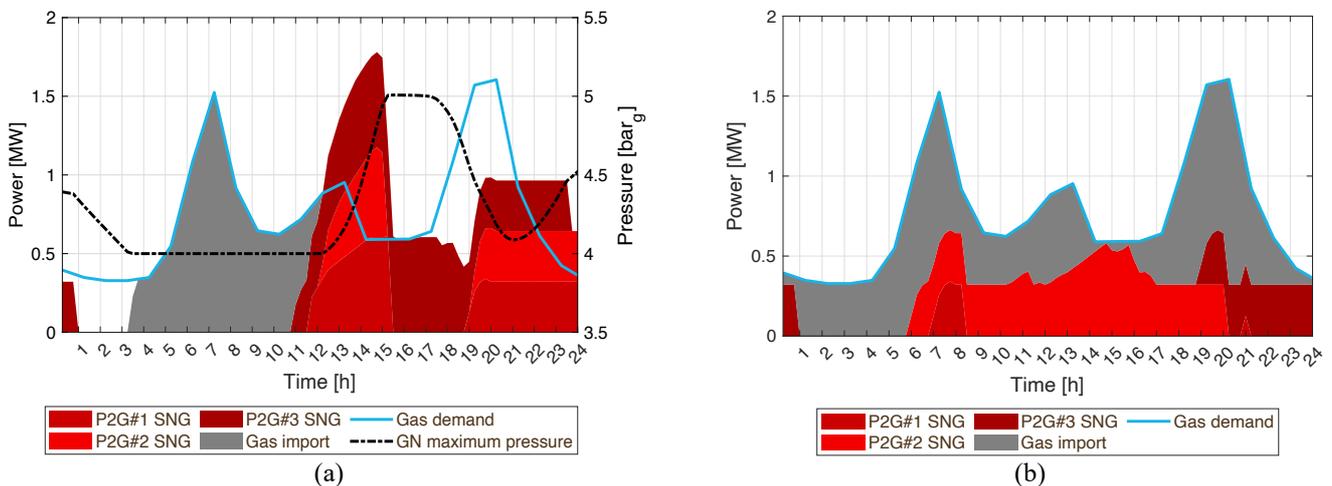

Figure 15. Gas network balance and pressure (summer day): (a) Reference case and (b) LPGN case.

---

[2] For the sake of clarity, the y-axis in Figure 14, Figure 15 and Figure 17 has different scales.

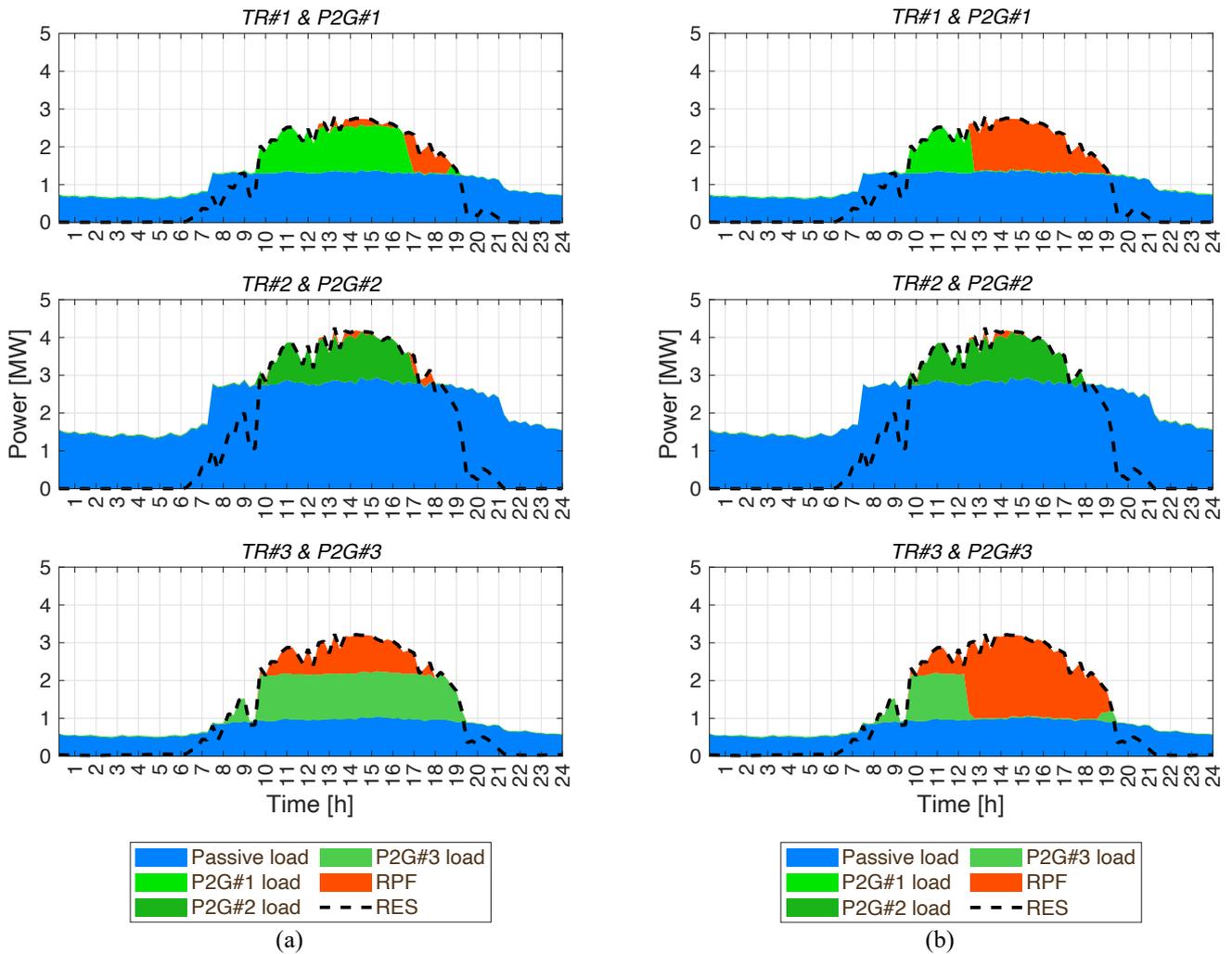

Figure 16. HV/MV transformer balance (summer day): (a) Reference case and (b) LPGN case.

4.3. Value of P2G modeling

The P2G lumped parameter model neglects the presence of the hydrogen buffer: in such an approach, the electricity energy is considered to be directly converted into SNG, without any possibility of storing hydrogen. Hence, the SNG production starts earlier than in the Reference case. Whenever the P2G reference model is used, the methanation units are only turned on when the hydrogen buffers have reached a predetermined state of charge (see Figure 17). It can be noted that the production of SNG in the LPP2G case is less uniform than in the reference case, as the lumped parameter model does not consider the ramp up and ramp down constraints of the plant, thus the SNG production follows the much faster variation characteristic of the electrolyzers. Apart from this misalignment, the use of this simplification during the heating season, when there is a high gas demand, does not substantially change the simulation results (see Table A. 1-3). The small differences that can be seen in Table A. 1, which are lower than 4 %, are mainly due to the fact that the reference model takes into account the change in energy conversion efficiency for different working conditions, while the lumped parameter model considers a constant average efficiency.

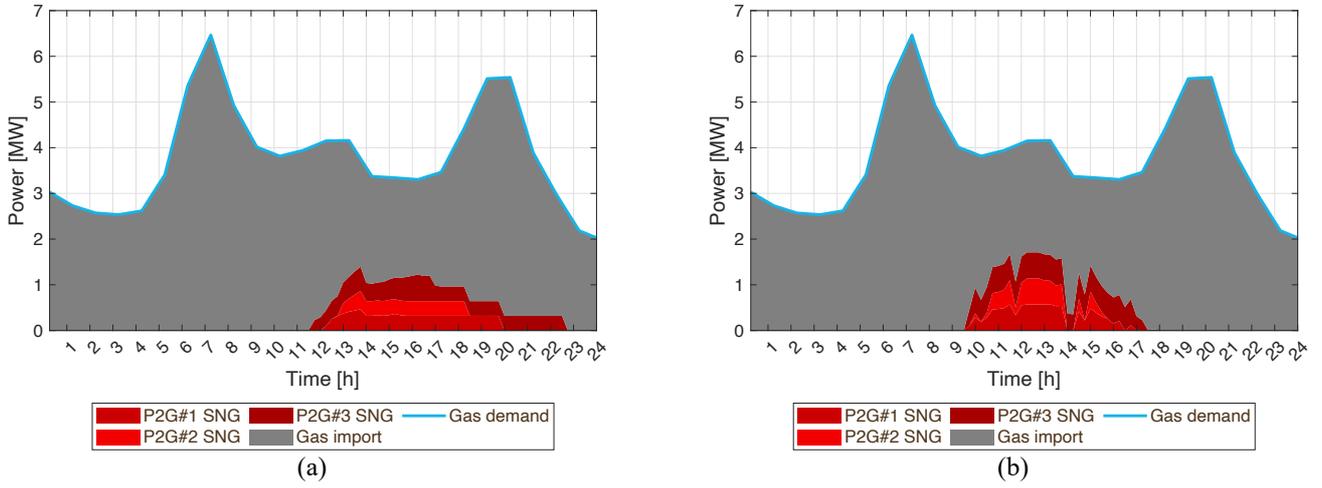
Figure 17. Gas network balance (Mid-season day): (a) Reference case and (b) LPP2G case.

However, in the summer season, due to the low gas demand, the gas network becomes less flexible, as it can accept a smaller quantity of SNG. In this case, neglecting the flexibility given by the decoupling of the methanation units from the electrolyzers affects the results of the simulation. In fact, when the gas network reaches the maximum allowed pressure, the electrolyzers could continue to work by accumulating the hydrogen produced inside the buffer. On the other hand, in the case of LPP2G, when the pressure in the gas network reaches its maximum limit at 12:45 (see Figure 18), the electrolyzers must also limit their loads (see Figure 19).

It should be noted that, even in the Reference case, the electrolyzer may be affected by restrictions (see P2G#1 in Figure 19a): this may happen when the hydrogen buffer reaches it maximum SoC, and the hydrogen production needs to be reduced to prevent an overpressure being created in the buffer. Nevertheless, without considering the hydrogen buffer, the P2G plants have less flexibility, which leads to the P2G potential being underestimated during the low gas demand period (i.e., in the summer season), and the use of P2G plants being underestimated by about 10%, which is also reflected by an equal underestimation of the SNG injection in the gas network. Moreover, the RPF on the electricity network is overestimated by about 150 % (see Table A. 1-3).

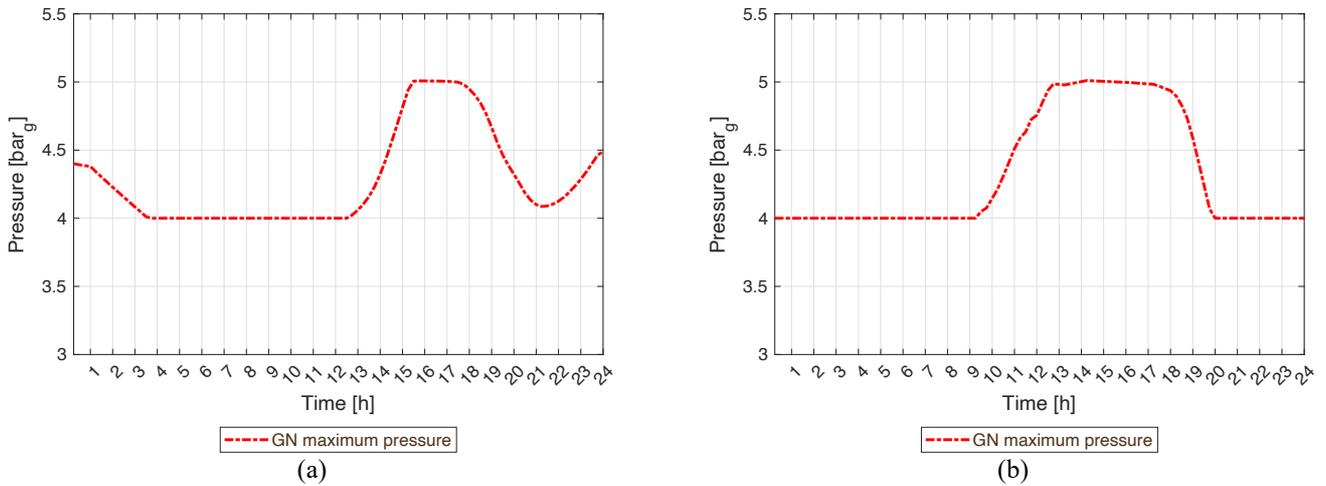
Figure 18. Gas network pressure (summer day): (a) Reference case and (b) LPP2G case.

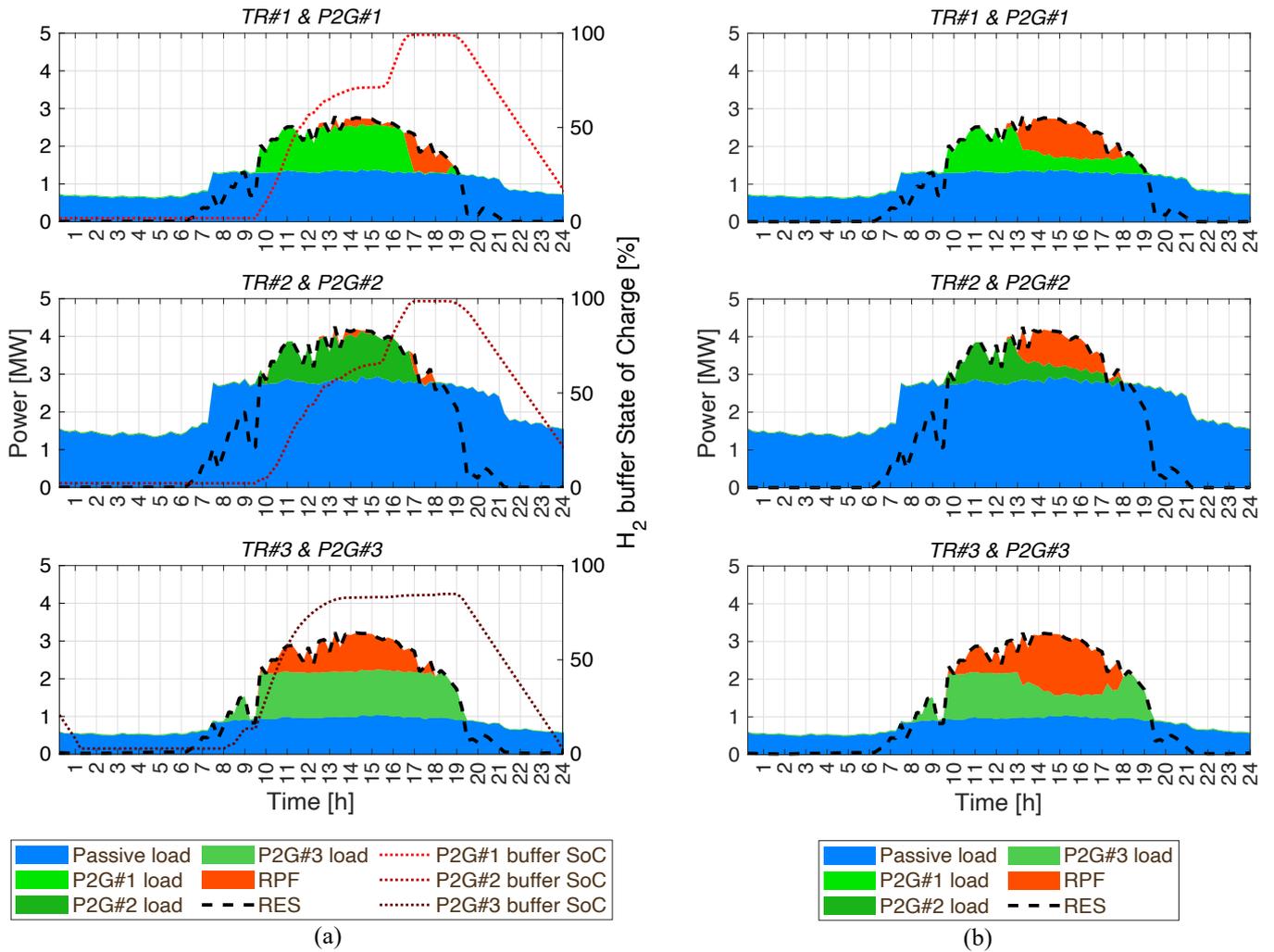

Figure 19. HV/MV transformer balance and hydrogen buffer SoC: (a) Reference case and (b) LPP2G case.

## 5. Conclusions

This paper has analyzed different modeling approaches for the simulation of multi-energy scenarios with electricity and gas distribution network connections to P2G plants. The scenario was specifically defined to create some critical situations for the use of these distributed resources in order to highlight whether, and under what conditions, the various models are able to correctly simulate the dynamics between the various components (the electricity network, the gas network and P2G plants).

The results obtained using detailed models of the three components have been compared with those obtained by simulating scenarios in which different modeling aspects had been neglected. Three simulations were carried out, in which the following aspects were neglected one at a time (*i*) the topology of the electricity distribution network and, consequently, the local power flows that occur within it, (*ii*) the topology of the gas network, the gas flows and the evolution of the network pressure and (*iii*) the intermediate energy conversion processes and storage lying in P2G plants between electricity and SNG.

The main conclusions that can be drawn from these analyses can be summarized as follows:

i. *Distribution System Modeling*: taking into account the topology of the electricity network makes it possible to evaluate in which area of the electricity network the over-generations of renewable energy occur and therefore allows one to choose the most appropriate resources to use, i.e., those closest to the electricity unbalances. Thus, the use of an electricity network model allows both the dispatching of flexible resources to be optimized and the best connection nodes for these resources to be evaluated.

- It is essential to consider electricity distribution network power flows if flexible resources are used to optimize the distribution network. Local VRES over-generations can only be highlighted by taking into account the distribution network topology. It is useful to analyze these phenomena as they can create RPFs on HV / MV transformers, thus leading to problems for the operation of the distribution system. Local over-generations of renewables can be mitigated, thanks to distributed flexible resources: thus, if these local imbalances are not detected, the potential benefit of using flexible distributed resources is underestimated.

- The single node representation of the electricity system cannot detect local overproductions. Nevertheless, it may be employed to evaluate whether it would be useful to offer the flexibility resources to TSO services for the operation of the transmission system.

ii. *Gas Grid Modeling*: simulating the dynamics of the gas network makes it possible more precisely evaluate the flexibility offered by this infrastructure, because the volume of the gas network can be used to store the production of SNG, thanks to the linepack effect. This storage is possible as long as the pressure in the network remains within the allowable pressure range. Hence, the gas network acts like a gas storage device, by allowing the gas demand to be decoupled from the SNG production.
- The gas network flexibility is relatively low in low gas demand periods: under certain conditions, the gas withdrawal can be lower than the SNG injection. When this happens, the gas network can reach saturation; thus, the SNG injections should be limited. Under these conditions, it is possible to evaluate the linepack potential and, hence, the gas network flexibility by considering the gas flows and the network pressure evolution. Since all the components in a multi-energy system scenario are closely connected, an underestimation of the gas network flexibility implies a lower flexibility of the connected P2G plants, which in turn leads to an underestimation of their utilization and also affects the operation of the electricity network (measured as residual RPF).
- The gas demand in a high gas demand scenario is normally much higher than for SNG injections. Such a high gas withdrawal makes the gas network flexibility higher, because the injected SNG may be absorbed directly by the user, without causing network saturation problems. In these cases, the linepack flexibility of the network may be neglected, without affecting the simulation results.

iii. *P2G Plant Modeling*: a physical model of a P2G plant allows all the processes that take place in a P2G plant to be simulated. This permits the decoupling between the methanation unit and the electrolyzer to be considered. The electrolyzer can therefore work even when, due to external restrictions, the methanation unit cannot operate. Taking these factors into account allows the flexibility of these plants to be properly estimated.
- On the one hand, in low gas demand periods, when the flexibility of the gas network is lower, the utilization of P2G plants may be constrained, because no SNG injection is allowed. In these circumstances, neglecting the interactions between the various components of the P2G system leads to an underestimation of the P2G flexibility and, therefore, an underestimation of both the possible SNG production and the services that these resources can offer to the electricity system.
- On the other hand, in high gas demand periods, the flexibility offered by the gas network is high enough to compensate for the underestimation of the P2G plant flexibility. In these conditions, the utilization of a simplified model that does not simulate the entire electricity-hydrogen-SNG chain in the P2G plant allows the energy flows within the multi-energy system, in which P2G plants are used to offer flexibility to the electricity network, to be evaluated with a good level of approximation.

Future works will concentrate on the use of these modeling hints to evaluate the possibility of locally managing multi-energy systems, that is, of implementing the concept of multi-energy microgrids. In such a case, the electricity network dynamics will take on a more important role, as will the consequent stress that the flexibility components have to sustain. Furthermore, new infrastructures, such as district heating and building storage, will be integrated by delineating a complete technical evaluation of the calculation tools that should be used for the design of future urban areas within the "smart cities paradigm".

## Appendix A. Simulation data results

Table A. 1. P2G plant results.

|  | Unit | Heating season | | | | Non-heating season | | | | Whole year | | | |
|---|---|---|---|---|---|---|---|---|---|---|---|---|---|
|  |  | P2G#1 | P2G#2 | P2G#3 | Tot | P2G#1 | P2G#2 | P2G#3 | Tot | P2G#1 | P2G#2 | P2G#3 | Tot |
|  |  | Reference case | | | | | | | | | | | |
| El. cons | GWh | 0.72 | 0.50 | 1.69 | 2.90 | 1.32 | 0.98 | 2.14 | 4.44 | 2.03 | 1.48 | 3.83 | 7.34 |
| SNG | GWh | 0.29 | 0.19 | 0.79 | 1.27 | 0.58 | 0.40 | 1.02 | 2.01 | 0.87 | 0.59 | 1.81 | 3.27 |
|  |  | Lumped parameter electricity network (LPEN) case | | | | | | | | | | | |
| El. cons | GWh | 0.80 | 0.80 | 0.80 | 2.40 | 1.44 | 1.43 | 1.40 | 4.26 | 2.23 | 2.23 | 2.20 | 6.66 |
| SNG | GWh | 0.33 | 0.33 | 0.33 | 1.00 | 0.64 | 0.64 | 0.62 | 1.91 | 0.98 | 0.97 | 0.96 | 2.90 |
|  |  | Lumped parameter gas network (LPGN) case | | | | | | | | | | | |
| El. cons | GWh | 0.71 | 0.50 | 1.68 | 2.90 | 0.85 | 0.81 | 1.56 | 3.23 | 1.57 | 1.31 | 3.25 | 6.12 |
| SNG | GWh | 0.29 | 0.19 | 0.79 | 1.27 | 0.35 | 0.34 | 0.72 | 1.42 | 0.64 | 0.53 | 1.51 | 2.68 |
|  |  | Lumped parameter P2G (LPP2G) case | | | | | | | | | | | |
| El. cons | GWh | 0.72 | 0.50 | 1.70 | 2.92 | 1.04 | 0.76 | 1.97 | 3.77 | 1.76 | 1.26 | 3.67 | 6.69 |
| SNG | GWh | 0.30 | 0.19 | 0.78 | 1.27 | 0.46 | 0.32 | 0.91 | 1.69 | 0.76 | 0.52 | 1.69 | 2.96 |

Table A. 2. Electricity network results.

|  | Unit | Heating season | | | | Non-heating season | | | | Whole year | | | |
|---|---|---|---|---|---|---|---|---|---|---|---|---|---|
|  |  | TR#1 | TR#2 | TR#3 | Tot | TR#1 | TR#2 | TR#3 | Tot | TR#1 | TR#2 | TR#3 | Tot |
| | | Reference case | | | | | | | | | | | |
| EL demand | GWh | 3.70 | 7.80 | 2.79 | 14.29 | 4.12 | 8.71 | 2.97 | 15.80 | 7.82 | 16.50 | 5.76 | 30.08 |
| RES | GWh | 2.41 | 3.23 | 4.12 | 9.76 | 3.90 | 5.68 | 5.39 | 14.97 | 6.32 | 8.91 | 9.51 | 24.73 |
| Surplus | GWh | 0.67 | 0.46 | 2.04 | 3.17 | 1.40 | 1.08 | 3.08 | 5.55 | 2.07 | 1.54 | 5.11 | 8.72 |
| Absorbed surplus | GWh | 0.62 | 0.40 | 1.61 | 2.63 | 1.23 | 0.90 | 2.07 | 4.20 | 1.86 | 1.30 | 3.68 | 6.83 |
| RPF | GWh | 0.05 | 0.06 | 0.43 | 0.53 | 0.16 | 0.18 | 1.01 | 1.35 | 0.21 | 0.24 | 1.44 | 1.88 |
| | | Lumped parameter electricity network (LPEN) case | | | | | | | | | | | |
| EL demand | GWh | - | - | - | 14.29 | - | - | - | 15.80 | - | - | - | 30.08 |
| RES | GWh | - | - | - | 9.76 | - | - | - | 14.97 | - | - | - | 24.73 |
| Surplus | GWh | - | - | - | 2.45 | - | - | - | 5.09 | - | - | - | 7.54 |
| Absorbed surplus | GWh | - | - | - | 1.98 | - | - | - | 3.80 | - | - | - | 5.78 |
| RPF | GWh | - | - | - | 0.46 | - | - | - | 1.29 | - | - | - | 1.76 |
| | | Lumped parameter gas network (LPGN) case | | | | | | | | | | | |
| EL demand | GWh | 3.70 | 7.80 | 2.79 | 14.29 | 4.12 | 8.71 | 2.97 | 15.80 | 7.82 | 16.50 | 5.76 | 30.08 |
| RES | GWh | 2.41 | 3.23 | 4.12 | 9.76 | 3.90 | 5.68 | 5.39 | 14.97 | 6.32 | 8.91 | 9.51 | 24.73 |
| Surplus | GWh | 0.67 | 0.46 | 2.04 | 3.17 | 1.40 | 1.08 | 3.08 | 5.55 | 2.07 | 1.54 | 5.11 | 8.72 |
| Absorbed surplus | GWh | 0.62 | 0.40 | 1.61 | 2.63 | 0.77 | 0.73 | 1.49 | 2.98 | 1.39 | 1.13 | 3.10 | 5.61 |
| RPF | GWh | 0.05 | 0.06 | 0.43 | 0.54 | 0.63 | 0.35 | 1.59 | 2.57 | 0.68 | 0.41 | 2.02 | 3.10 |
| | | Lumped parameter P2G (LPP2G) case | | | | | | | | | | | |
| EL demand | GWh | 3.65 | 7.74 | 2.74 | 14.13 | 4.07 | 8.66 | 2.91 | 15.64 | 7.72 | 16.40 | 5.65 | 29.77 |
| RES | GWh | 2.41 | 3.23 | 4.12 | 9.76 | 3.90 | 5.68 | 5.39 | 14.97 | 6.32 | 8.91 | 9.51 | 24.73 |
| Surplus | GWh | 0.68 | 0.47 | 2.06 | 3.21 | 1.42 | 1.09 | 3.10 | 5.61 | 2.10 | 1.56 | 5.17 | 8.82 |
| Absorbed surplus | GWh | 0.64 | 0.41 | 1.65 | 2.70 | 0.98 | 0.68 | 1.92 | 3.58 | 1.61 | 1.10 | 3.57 | 6.28 |
| RPF | GWh | 0.04 | 0.06 | 0.42 | 0.51 | 0.44 | 0.41 | 1.18 | 2.03 | 0.48 | 0.46 | 1.60 | 2.55 |

Table A. 3. Gas network results.

|  | Unit | Heating season | Non-heat. season | Whole year |
|---|---|---|---|---|
| | | Reference case | | |
| Gas demand | GWh | 31.65 | 4.37 | 36.02 |
| Imported Gas | GWh | 30.38 | 2.37 | 32.75 |
| Imported Gas | GWh | 1.27 | 2.00 | 3.27 |
| | | Lumped parameter electricity network (LPEN) case | | |
| Gas demand | GWh | 31.65 | 4.37 | 36.02 |
| Imported Gas | GWh | 30.65 | 2.47 | 33.12 |
| SNG | GWh | 1.00 | 1.91 | 2.90 |
| | | Lumped parameter gas network (LPGN) case | | |
| Gas demand | GWh | 31.65 | 4.37 | 36.02 |
| Imported Gas | GWh | 30.38 | 2.96 | 33.34 |
| SNG | GWh | 1.27 | 1.42 | 2.68 |
| | | Lumped parameter P2G (LPP2G) case | | |
| Gas demand | GWh | 31.65 | 4.37 | 36.02 |
| Imported Gas | GWh | 30.38 | 2.68 | 33.06 |
| SNG | GWh | 1.27 | 1.69 | 2.96 |